\documentclass[twocolumn,secnumarabic,amssymb,amsmath,nofootinbib,tightenlines,showpacs,nobibnotes,aps,pra]{revtex4}
\usepackage{amsfonts}
\usepackage{bm}%
\usepackage{slashed}
\usepackage{subfigure}%
\usepackage[colorlinks=true,linkcolor=blue]{hyperref}%
\usepackage{graphicx}
\usepackage{dcolumn}

\begin{document}

\title{Quantum Spin Hall Effect as $\mathbb{Z}_2$ Global Gauge Anomaly}
\author{Yunqin Zheng}
\author{Shaolong Wan}
\altaffiliation{Corresponding author}
\email{slwan@ustc.edu.cn}
\affiliation{Institute for Theoretical
Physics and Department of Modern Physics University of Science and
Technology of China, Hefei, 230026, \textbf{P. R. China}}

\date{\today}

\begin{abstract}
We study the relation between the quantum spin Hall effect(QSHE) and the
$\mathbb{Z}_2$ global gauge anomaly and discover that there exists an
one-to-one correspondence between them. By constructing a two
dimensional non-abelian gauge theory whose non-abelian gauge field
is the Berry connection induced by the Bloch wave function of the
quantum spin Hall system, we prove that if the quantum spin Hall
system is topologically nontrivial, the corresponding 2D gauge
theory has a global gauge anomaly. We further generalize our
discussion of the zero modes of the Dirac operator which play a
central role in our analysis. We find that the classical "spin"
Hall effect also contains a $\mathbb{Z}_2$ topological invariant
which is not noticed before as far as we know.
\end{abstract}


\maketitle

\section{Introduction}

Topological insulator (TI) discovered by C. L. Kane and E. J.
Mele\cite{Kane} is a new state of matter whose bulk behaves like
an insulator while its edge behaves like a metal. The mechanism
under-lying the phenomenon is the time-reversal symmetry. The
edge, where the energy gap is closed, is robust against
perturbations as long as the time-reversal symmetry is preserved.
This is the reason why it is named a
symmetry-protected-topological(SPT) state more precisely. Each TI
system is labelled by a topological invariant, which indicates its
topological properties. Transition between states with different
topological invariants happens only if the energy gap closes. The
reader can refer to Refs. \cite{Qi-Zhang, Qi-Zhang2, Hasan-Kane}
and references therein for a complete review.

Now, many progress on the topological insulators have been made,
among which the discovery of the $\mathbb{Z}_2$ topological
property of the quantum spin Hall states is a landmark. This
$\mathbb{Z}_2$ topological property differs greatly  from those
ordinary quantum Hall effects which are $\mathbb{Z}$ classified.
It states that in the Brillouin zone, there can stably exist only
one or zero Dirac-cone-like point which will be defined more
precisely below. The presence of more than one such point is
non-stable, implying that we can perturb the parameters so that
they can be annihilated and created in pairs, until there are only
one pair or none remains in the Brillouin zone. Then, the authors
of \cite{QHZ} proposed another formalism to describe the quantum
spin Hall effects, and made a generalization to high dimensional
topological insulators using a Chern-Simons field theory. The
method is mainly dimensional reduction, and they acquired a
seemingly different set of topological invariants.
Wang\cite{Z.Wang} proved that in three dimensions, the two set of
topological invariants are equivalent, which unified the two
different approaches.

All above achievements are based on non-interacting systems, and
the band structure analysis is valid. However, when interaction is
taken into consideration, many of the above discussion fail. Some
results have been acquired by using Green function\cite{Wang,
Wang1, Wang2}. But much more are still unknown. Another way to
understand the interacting phenomena is to study anomalies related
to topological states since anomalies are not restricted to
non-interacting systems. It has long been believed that various
anomalies and topological phases are in one-to-one correspondence.
In Ref.\cite{Ryu}, the authors made a summary of the
correspondence between $\mathbb{Z}$ topological states and
anomalies, but the situation for most $\mathbb{Z}_2$ states is
unclear. This is our primary motivation in this work and we
propose a relationship between the quantum spin Hall effect (QSHE)
$\mathbb{Z}_2$ invariant and the global gauge anomaly.

The article is organized as follows. In Sec.\ref{sec2}, we give a
brief review according to Kane-Mele's work\cite{Kane-Mele} on a
$\mathbb{Z}_2$ topological invariant. In Sec.\ref{sec3}, we
summarize some main features of the global gauge anomaly which are
relevant to our discussion. In Sec.\ref{sec4}, the relationship
between the gauge anomaly and the $\mathbb{Z}_2$ topological
invariant is established. As an application of our formalism, in
Sec.\ref{sec5}, we discuss the Morse theory associated to the
Dirac operator and find a bonus, which indicates that the
classical "spin" Hall effect also have a $\mathbb{Z}_2$ internal
topological invariant as in the quantum case predicted nearly a
decade ago. Finally, we give some discussion and conclusion, and
propose some further topics to be investigated in Sec.\ref{sec6}.

\section{Topological $\mathbb{Z}_2$ invariants of a $2+1$ dimensional quantum spin Hall
system}
\label{sec2}

Before studying the relation between the quantum spin hall effect and the
global gauge anomaly, it is beneficial to give a brief review on
both theories.

According to the work by C. L. Kane and E.J. Mele\cite{Kane-Mele},
we first consider a two spatial dimensional system, without
interaction, with a Hamiltonian $H(k)$. The corresponding Bloch wave
functions $|k, \alpha \rangle$ are given by the eigen-equation
\begin{equation}
H(k) |k, \alpha \rangle = E_{\alpha} |k, \alpha \rangle.
\label{2.1 H}
\end{equation}
We require the Hamiltonian to be the time-reversal invariant:
$\Theta H(k) \Theta^{\dagger} = H(-k)$, where the time-reversal
operator is $\Theta = T K$ and $K$ is a complex conjugate
operator. Due to such a symmetry, both $|-k,\alpha \rangle$ and
$\Theta |k, \alpha \rangle$ are solutions of an eigen-equation
with an eigenvalue $E_{\alpha}$. They may be related by fixing the
gauge condition as $|-k, \alpha \rangle = \Theta |k, \alpha
\rangle$.

In order to define topological invariants, it is convenient to
introduce two concepts: an odd space and an even space. The even space
is a set of points in momentum space satisfying the condition:
$\Theta H(k) \Theta^{\dagger} = H(-k) = H(k)$, which means that
$\Theta|k,\alpha\rangle$ and $|k,\beta\rangle$ are not
perpendicular to each other, while the odd space is a set that the
two vectors are mutually orthonormal. The two space can be
distinguished by considering whether the matrix $\langle k, \alpha
| \Theta |k, \beta \rangle$ equals to zero(an odd space) or not(an even
space).

Now, we list some properties of the matrix $\langle k, \alpha |
\Theta |k, \beta \rangle$:
\begin{enumerate}
\item It is anti-symmetric, $i.e.~ \langle
k,\alpha|\Theta|k,\beta\rangle=-\langle
k,\beta|\Theta|k,\alpha\rangle$. And we can define its Pfaffian as
$P(k) = Pf(\langle k,\alpha|\Theta|k,\beta\rangle)$.

\item The points in an odd space appears in pairs at $(k,-k)$.

\item Since points in an odd space are in pairs, when there are
two such pairs in the Brillouin zone,   it is always possible to
the move them together to annihilate them.
\end{enumerate}
From the above discussion, we know that the number of pairs in an odd space can
only be zero or one. The two cases cannot be deformed to each
other since if $k$ wants to meet $-k$, the only choice is to meet
at the origin, which obviously belongs to an even space. Therefore,
there is a topological obstruction between the two phases. So, it is
possible to define the number of pairs in an odd space as a
topological invariant which distinguishes the two cases. In order
to calculate it, it is convenient to express it as a contour
integral with the help of the Cauchy theorem:
\begin{eqnarray}
I=\frac{1}{2\pi i}\oint_{half BZ}d\vec{k}\cdot
\nabla_{\vec{k}}\log (P(k)+i\delta)~ (mod\ 2). \label{2.2 I}
\end{eqnarray}
Integrating over half of the Brillouin zone means that we only have to
count points instead of pairs with $\delta$ being a small parameter
making the integral convergence.

In the above discussion, we only require the Hamiltonian to be
time-reversal symmetric, and do not impose any other restrictions,
such as a rotational symmetry or the specific form of the
Hamiltonian. So the $\mathbb{Z}_2$ topological properties are quite universal.

\section{A brief review of the global gauge anomaly}
\label{sec3}

The global gauge anomaly, especially an $SU(2)$ gauge anomaly, was
originally discovered by E. Witten\cite{Witten} by considering a
chiral Weyl fermion minimally coupled to an $SU(2)$ gauge field.
The action of the model is
\begin{eqnarray}
S = \int d^4 x \left[\bar{\psi} i \slashed{D} \psi - \frac{1}{2
g^2} \textrm{tr} F_{\mu\nu} F^{\mu\nu} \right] \label{3.1 S}
\end{eqnarray}
where $\psi(x)$ is a Weyl fermion, $A_{\mu}$ is an $SU(2)$ gauge
field, $D_{\mu} =
\partial_{\mu} + i g A_{\mu}$ is a covariant derivative and
$F_{\mu\nu}=\partial_{\mu}A_{\nu}-\partial_{\nu}A_{\mu}-ig[A_{\mu},A_{\nu}]$
is the field strength. In order to quantize the theory, we consider the
partition function:
\begin{eqnarray}
Z&=&\int d\psi d\bar{\psi}\int dA_{\mu}\nonumber\\&&\times\exp \left\{-\int d^4x
\left[\bar{\psi}i\slashed{D}\psi-\frac{1}{2g^2}trF_{\mu\nu}F^{\mu\nu}\right]
\right\}. \label{3.2 Z}
\end{eqnarray}
Integrating over the fermion degree of freedom, we may get an
effective low energy action only in terms of the gauge fields. To
do the integration, we consider the case for a Dirac fermion
coupled to an $SU(2)$ gauge field. For this case, the integration
is well known to be $\det(i\slashed{D})$. Since the number of Weyl
spinors is half of the number of Dirac spinors, thus, the
integration is $\pm(\det(i\slashed{D}))^{\frac{1}{2}}$. Notice
that the sign at the front cannot be fixed globally. In order to
settle this ambiguity, one needs to choose a specific
configuration, say, the vacuum $A^0_{\mu}$, to be positive and
other configurations related to this via a gauge transformation
$A_{\mu}=U^{-1}A^0_{\mu}U-iU^{-1}\partial_{\mu}U$, where $U\in
SU(2)$.

In four dimensions, more generally for any even dimension, aside
from all the Gamma matrices, there is an additional one $\Gamma_5$
which is proportional to the product of all Gamma matrices and
anti-commute with all others. Thus, when $\psi$ is an eigenstate
with energy $E$, $\Gamma_5\psi$ is also an eigenstate with the
opposite energy $-E$. Therefore, the energy spectrum is symmetric
with respect to the origin. Considering that $\det(i\slashed{D})$
is just the product of all eigenvalues,
$\pm(\det(i\slashed{D}))^{\frac{1}{2}}$ is determined by a product
of half of the spectrum, picking one value from each pair. In
order to let the sign in the front to be definitive, one needs to
exclude all the zero eigenvalues from the definition of the
determinant.

From the above discussion, one can find that there exists some $U$
such that $(\det i\slashed{D}[A])^{\frac{1}{2}}=-(\det
i\slashed{D}[A^U])^{\frac{1}{2}}$ and others $(\det
i\slashed{D}[A])^{\frac{1}{2}}=(\det
i\slashed{D}[A^U])^{\frac{1}{2}}$, which is known as a global
gauge anomaly. More detailed discussions show that the measures of
$U$ in both cases turn out to be the same. The consequence of such
a gauge anomaly can lead to the above partition mathematically
ill-defined. But we do not spare our efforts on it. As Witten
pointed out, the mathematical origin of a global gauge anomaly
lies in the fact that
\begin{equation}
\pi_4(SU(2))=\mathbb{Z}_2, \label{3.3 Z2}
\end{equation}
where the subindex $4$ is the dimension of space-time. Equation
(\ref{3.3 Z2}) shows the appearance of an anomaly is determined by
the topological property of the gauge group. Indeed, when we
deform the transformation matrix a little, we can prove that the
resulting gauge field belongs to the same class as the original
one. The result also tells us that the global gauge anomaly can
occur in other even dimensions since Clifford algebra can have
symmetric energy spectrum only in even dimensions and its square
root can have a definitive meaning\cite{Nakahara}.

In the next section, we will in fact use a more general case
\begin{equation}
\pi_2(G)=\mathbb{Z}_2 \label{3.4 Z2}
\end{equation}
where the subindex $2$ is the dimension of momentum space and $G$
is the symmetry group of a QSHE Hamiltonian. It is worth to note
that Eq.(\ref{3.4 Z2}) is true and can be proven almost the same
as for the four dimensional case.

\section{$\mathbb{Z}_2$ topological invariant in a QSHE from a global gauge anomaly}
\label{sec4}

The similarities of the topological property between the QSH and
the gauge anomaly makes us to ask whether there exists a relation
between them and what it is. Indeed, by studying, we find that the
system described by a two dimensional Lagrangian which carries the
topological information of the QSH system exhibits a global gauge
anomaly. In the following, we show how they are obtained.

Firstly, we display some correspondence between a QSHE and a global
gauge anomaly:
\begin{itemize}
  \item \textbf{Field contents}: Bloch wavefunction $|k,\alpha\rangle$ or
  Pfaffian$(\langle k,\alpha|\Theta|k,\beta\rangle)$ $\longleftrightarrow$ Nonabelian
  gauge field $A_{\mu}$.
  \item \textbf{Base Manifold}: 2D momentum space $\longleftrightarrow$ Even dimensional D
  space-time.
  \item \textbf{Symmetry}: Time Reversal Symmetry $\longleftrightarrow$ Chiral-Antichrial
  Symmetry.
  \item \textbf{Topological origin}: $\pi_2(G)=\mathbb{Z}_2$,
  where $H\in G$ $\longleftrightarrow$ $\pi_D(G)=\mathbb{Z}_2$, where
  $A_{\mu} \in G$.
\end{itemize}
Although there appears to be some difference, the similar
topological origin indicates both are determined by the
homotopic group of some group which the field content belongs to.
This motivates us to find a connection between them. In order to
connect the non-abelian gauge field with the Bloch wavefunction,
the most natural way is to define the Berry connection/gauge field
as follows:
\begin{equation}
A_{\mu}^{\alpha\beta}(k)=-i\langle
k,\alpha|\partial_{\mu}|k,\beta\rangle. \label{4.1 GF}
\end{equation}
Notice that it is indeed a gauge field because when the Hamiltonian
takes an unitary transformation as $H(k) \rightarrow U H(k)
U^{-1}$, the Bloch wavefunction is transformed by $|k, \alpha
\rangle \rightarrow U_{\alpha\beta}|k, \beta \rangle$ and the
Berry connection is transformed by $A_{\mu}^{\alpha\beta}(k)
\rightarrow
(U^{*}A_{\mu}U^{*\dagger}-iU^{*}\partial_{\mu}U^{*\dagger})^{\alpha\beta}$,
which indicates that $A_{\mu}$ transforms similarily as a gauge field.
From the above argument, we may find that the representation of
the Hamiltonian and that of the Berry connection belong to the
same group. Therefore, the topological origin of both theories
exactly matches --- both of them are determined by the symmetry
group which the non-abelian gauge field
belongs to.

Although the origin is the same, it is still worth to study whether
there is a one-to-one correspondence between a topological
trivialness/nontrivialness and a gauge anomaly appearance/vanishness. So further
investigations are needed.

As we know, for any antisymmetric matrix, $[Pf(X)]^2=\det(X)$, so
$Pf(X)=\pm[\det(X)]^{\frac{1}{2}}$, where the sign $\pm$ is
indefinitive. In order to understand the role which the Berry
connection is playing in the $\mathbb{Z}_2$ topological invariant,
we substitute it into Eq.(\ref{2.2 I}), the invariant can be given
as
\begin{widetext}
\begin{eqnarray}
I&=&\frac{1}{4\pi i}\oint_{half BZ}d\vec{k}\cdot\nabla_{\vec{k}}\log (\det(\langle k,\alpha|\Theta|k,\beta\rangle)) (mod\ 2)\nonumber\\
&=&\frac{1}{4\pi i}\oint_{half BZ}d\vec{k} tr([^*\langle
k,\beta|T^{\dagger}\partial_{\vec{k}}|k,\alpha\rangle]^*+\langle
k,\alpha|\partial_{\vec{k}}T|k,\beta\rangle^*)[\frac{1}{\langle
k,\alpha|T|k,\beta\rangle^*}](mod\ 2).\nonumber\\
\label{4.2 II}
\end{eqnarray}
\end{widetext}
Here the indefinitive sign disappears due to the $\nabla$ operator.
The first term in the bracket can be rewritten as
\begin{eqnarray}
[^*\langle
k,\beta|T^{\dagger}\partial_{\vec{k}}|k,\alpha\rangle]^* =
iS^{\beta\gamma}(k)[\vec{A}^{\gamma\alpha}(k)]^*, \label{4.3}
\end{eqnarray}
where we define $S^{\alpha\beta}=\langle
k,\alpha|\Theta|k,\beta\rangle$.

As for the second term, some special care is needed. From the second
section, we have chosen the gauge fixing condition as
$|-k,\alpha\rangle=T|k,\alpha\rangle^*$ and to expand it in terms
of gauge field, we need to use this identity. One should pay
attention to the fact that the identity can only be defined
locally on the base manifold, otherwise, as one can check without
much effort, the contributions of the first term and the second
term just cancel with each other, indicating that the QSHE system is always
topological trivial. We have to remark that the argument in the
second section does not solve this problem since the equation only
applies to the proof of the orthonormal condition, which is a purely
local argument. But in this case, we have to integrate over the
Brillouin zone and thus a global effect emerges. The simplest
solution to tackle the problem is to divide the Brillouin
zone into two parts, namely $A$ and $B$. In part $A$, the above
condition is valid, while in part $B$, the Bloch wave function is
related to that of $A$ via a transformation matrix $t$:
$|k,\alpha\rangle_A=t^{\alpha\beta}|k,\beta\rangle_B$. When $k$ is
in part $A$, the second term can be given as
\begin{eqnarray}
_A \langle k, \alpha | \partial_{\vec{k}} T |k, \beta \rangle^*_A
= -i S^{\alpha\gamma}_A (k) \vec{A}^{\gamma\beta}_A (-k).
\label{4.4}
\end{eqnarray}
When $k$ is in part B, the second term can be expressed as
\begin{eqnarray}
_B\langle k,\alpha|\partial_{\vec{k}}T|k,\beta\rangle^*_B &=&
t^{*\delta\gamma}S_B^{\alpha\gamma}(\partial_{\vec{k}}t^{\delta\beta})
-t^{\delta\beta}\partial_{\vec{k}}t^{*\delta\gamma}S_B^{\alpha\gamma}\nonumber\\
&&-it^{\delta\beta}t^{*\delta\gamma}S_B^{\alpha\eta}\vec{A}_B^{*\eta\gamma}(k).
\label{4.5}
\end{eqnarray}
Substituting all Eqs.(\ref{4.3})-(\ref{4.5}) into Eq.(\ref{4.2
II}), we obtain the topological invariant as follow
\begin{eqnarray}
I&=&\frac{1}{2\pi i}\oint_{\partial B}d\vec{k} \textrm{tr} (t^{*\delta\gamma}\partial_{\vec{k}}t^{\delta\beta})(mod\ 2)\nonumber\\
&=&\frac{1}{2\pi i}\oint_{\partial B}d\vec{k} \textrm{tr}
(t^{\dagger}\partial_{\vec{k}}t)(mod\ 2). \label{4.6 I}
\end{eqnarray}
Here, its physical meaning is obvious: when momentum $k$ travels
around the part $B$ once, $t$ gains an additional phase angle,
which is integer times $2\pi i$. Notice that $B$ in this case may not be simply
connected, but it will not affect our analysis because we can add
different branches of integral provided that we carefully keep the same
orientation among them. The topological invariant just
counts this integer. Moreover, when we set $A_{\mu}^0$ to be a
trivial Berry connection, and let $t$ be the gauge transformation,
we get a new Berry gauge field. The invariant can be expressed in
terms of the new gauge field via
\begin{equation}
I=\frac{1}{2\pi}\oint_{BZ} \textrm{tr}[A](mod\ 2) \label{4.7 I}.
\end{equation}
We have to remark here that this result was previously obtained by
L. Fu and C. Kane\cite{Fu-Kane} from a more complicated
method. Here we acquire it by a simple and direct calculation. With
the help of this simplified invariant, we can state an important
property which directly leads to the equivalence between the two
theories topologically.

If the two Hamiltonians $H_1(k)$ and $H_2(k)$ describe QSHE
systems, they determine two Berry connections $A_1(k)$ and
$A_2(k)$, as well as two transformation matrices $t_1$ and $t_2$.
They correspond to two topological invariants
\begin{eqnarray}
I_1=\frac{1}{2\pi i}\oint_{\partial B}d\vec{k} \textrm{tr} (t^{\dagger}_1\partial_{\vec{k}}t_1)(mod\ 2)=0\nonumber\\
I_2=\frac{1}{2\pi i}\oint_{\partial B}d\vec{k} \textrm{tr}
(t^{\dagger}_2\partial_{\vec{k}}t_2)(mod\ 2)=1. \label{4.8 II}
\end{eqnarray}
We define a new Berry connection connecting the above two
adiabatically as $A_{s\mu}=(1-s)A_{1\mu}+sA_{2\mu}$ and
$s\in[0,1]$. All the above are based on the information provided by the
QSHE. Now, we define a two dimensional theory whose Lagrangian is
\begin{equation}
\mathcal{L}=\bar{\psi}i\slashed{D}\psi-\frac{1}{2e^2}\textrm{tr}F_{\mu\nu}F^{\mu\nu},
\label{4.9 L}
\end{equation}
$F_{\mu\nu}$ is the strength field induced by the Berry gauge
field and $\psi$ is a Weyl spinor. The Dirac equation, defined on
the momentum space, can be written as
\begin{equation}
i\slashed{D}[A_s]\psi_s=\lambda_s\psi_s. \label{4.10 Deq}
\end{equation}
As the $s$ runs adiabatically from $0$ to $1$, we can follow the
tracks of the eigenvalue $\lambda$. In the following, we prove that if
there exist odd number of eigen-spectral lines flowing across zero,
they lead to the fact that
\begin{eqnarray}
(\det i\slashed{D}'[A_1])^{\frac{1}{2}}=-(\det
i\slashed{D}'[A_2])^{\frac{1}{2}}, \label{4.11}
\end{eqnarray}
where the prime on Dirac operators means they are induced by the
Dirac fermions, not the Weyl fermions. Therefore, there is a global
gauge anomaly in the system defined by Lagrangian (\ref{4.9 L}).

Here, we need to make a few remarks that the Dirac equation
(\ref{4.10 Deq}) and the Lagrangian (\ref{4.9 L}) stated above are
slightly different from normal ones, which are defined on the
momentum space.

\begin{figure}[!ht]
\centering
\includegraphics[width=0.22 \textwidth]{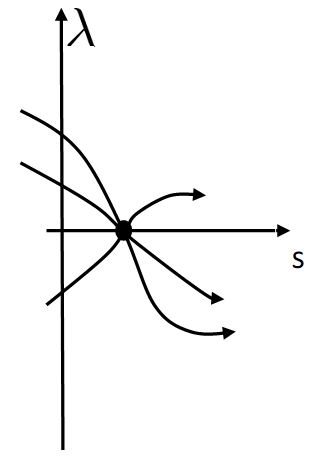}
\caption{Zero modes of Dirac operator at $s$. This zero point
consists of three spectral flows: two from the above and one from
below, i.e. $\nu_+=2$ and $\nu_-=1$¡£} \label{fig1}
\end{figure}

Since the topological invariant is $\mathbb{Z}_2$ valued, any even
number can be adiabatically deformed to zero; similarly, any odd number
to one. Thus we may remove the $(mod\ 2)$ operator and just set
the integrals to be one and zero. What is more, since $H_1(k)$ is
topological trivial, we may set the corresponding Berry connection
and the transformation matrix identically as zero. Therefore, we
may set the gauge transformation relating $A_1$ and $A_2$ as
$t\equiv t_2$, $i.e.$ $A_{2\mu} = t^*A_{1\mu}t^{*\dagger} -
it^*\partial_{\mu}t^{*\dagger}$. For the two dimensional case,
according to the Atiyah-Singer index theorem\cite{Friedan, Vafa}
\begin{eqnarray}
\nu_+ - \nu_-=\frac{1}{2\pi}\int_{BZ}
\textrm{tr}F=\frac{1}{2\pi}\oint_{BZ} \textrm{tr}A, \label{4.12
index th}
\end{eqnarray}
we can obtain
\begin{eqnarray}
\nu^s_+ -\nu^s_- &=&\frac{s}{2\pi}\oint_{\partial BZ}d\vec{k}
\textrm{tr} (t^{\dagger}\partial_{\vec{k}}t)\nonumber\\
&=&\frac{s}{\pi}\oint_{\partial B}d\vec{k} \textrm{tr} (t^{\dagger}\partial_{\vec{k}}t) \nonumber\\
&=&2sI_2=2s , \label{4.13 index}
\end{eqnarray}
where in the second line, we switch the integration domain only in
part $B$. On the left hand side, $\nu_+$ counts the number of zero
modes of the dimensional kernel of the left hand Dirac operator $iP_+
\slashed{D}$ where $P_+ =\frac{1}{2} (1 +
\gamma^{2+1})$\cite{Nakahara}. Similar definition works for
$\nu_-$. Here, we will use a more convenient and straight forward
way to compute the index with the help of Fig.\ref{fig1}.
$\nu_+$ represents the number of flows from up to down and $\nu_-$
from down to up. Therefore, the left hand side must be an integer.
As for the right hand side, $s\in [0,1]$, it means that $s$ can only
be $0$ or $\frac{1}{2}$ or $1$. When $I_2 = - 1$, $s$ reverse its
sign. We claim that $s=\frac{1}{2}$ is the only solution in the
present case because $s=0$ or $s=1$ implies that there are zero modes
in the original spectrum, inconsistent with our assumption.
Therefore, when $s$ goes from $0$ to $1$, the only possible point
for zero mode is at $s=\frac{1}{2}$.

\begin{figure}[!ht]
\centering
\includegraphics[width=0.25\textwidth]{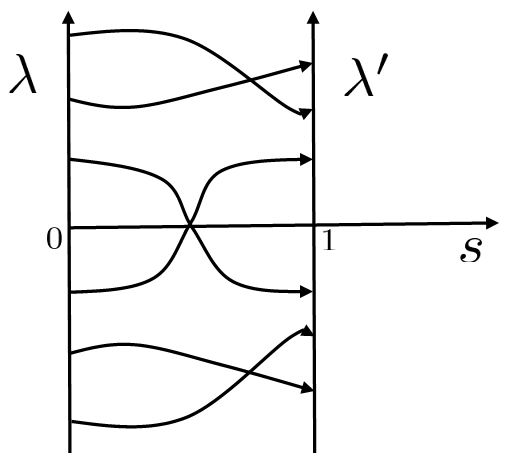}
\caption{A typical flowing spectrum. The horizontal axis
represents parameter varying from zero to one. The vertical axis
represents eigenvalues. Eigenvalues run as the parameter $s$
vary.} \label{fig2}
\end{figure}
\begin{figure}[!ht]
\centering
\includegraphics[width=0.25\textwidth]{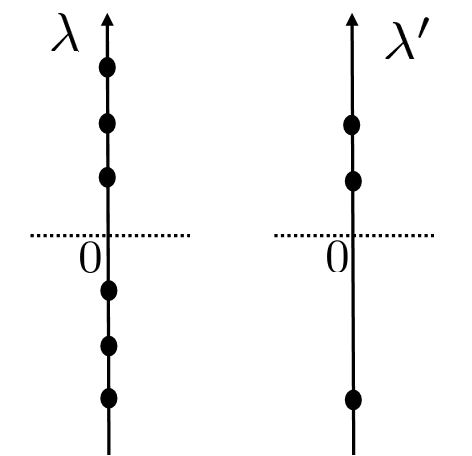}
\caption{An example of Dirac and Weyl spectrum. The left is Dirac
and the right is Weyl.} \label{fig3}
\end{figure}

Fig.\ref{fig2} is a typical flowing spectrum of that Dirac operator.
The spectrum is symmetric with respect to the zero value, as
previously stated. As $s$ varies from 0 to 1, the eigenvalues
rearrange to form a new spectrum. Although the overall
distribution seems unchanged, the relative position may vary. In
order to acquire the spectrum of an operator induced by a Weyl
fermion, we can choose one eigenvalue from each pair.
Fig.\ref{fig3} are a pair of Dirac and Weyl spectrum.
\begin{figure}[!ht]
\centering
\includegraphics[width=0.35\textwidth]{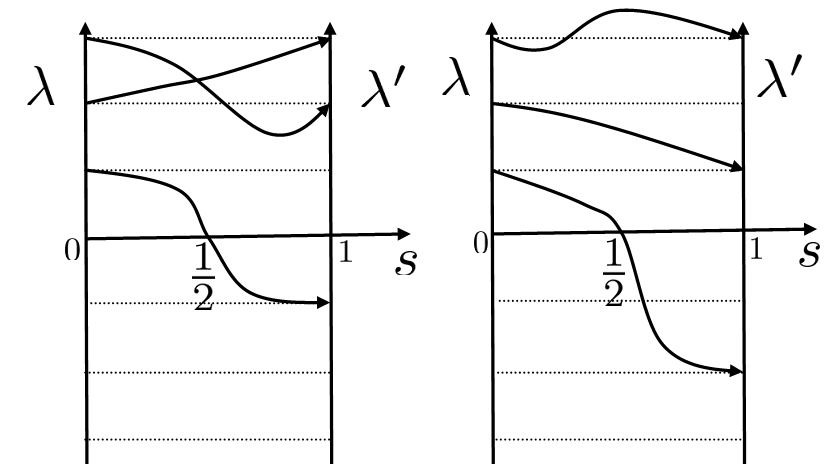}
\caption{When there is only one eigenvalue running across the
horizontal axis, the product of all the eigenvalues reverse sign.
Here we only list two possibilities. } \label{fig4}
\end{figure}

Now, we search for all the solutions satisfying $\nu_+ -\nu_- = 1$
and prove that all these solutions are with a global gauge
anomaly.

First, when there is one spectral line running across horizontal
axis, as shown in Fig.\ref{fig4}. In order to guarantee the
spectrum to be Weyl, the only possibilities are those which satisfies
 that the product of all the eigenvalues being the same absolute value
of the original ones. Moreover, since there is only one eigenvalue
which reverses its sign, though its absolute value may change as
in Fig.\ref{fig2}, the product is exactly the minus of the
Original one, which is equivalent to the identity
\begin{eqnarray}
(\det i\slashed{D}'[A_1])^{\frac{1}{2}}=-(\det
i\slashed{D}'[A_2])^{\frac{1}{2}}. \label{4.14}
\end{eqnarray}
So, for this case, we prove that there is a global gauge anomaly.

Second, when there are two eigenvalues flowing across the
horizontal axis, the two can both go from up to down, which means
$\nu_+ - \nu_- = 2$, or in the opposite direction, i.e. $\nu_+ -
\nu_- = -2$. Another possibility is that one from up to down and the
other from down to up, i.e. $\nu_+ - \nu_- =0$. All the above
three cases do not match the condition $\nu_+ - \nu_- = 1$.

\begin{figure}[!ht]
\centering
\includegraphics[width=0.24\textwidth]{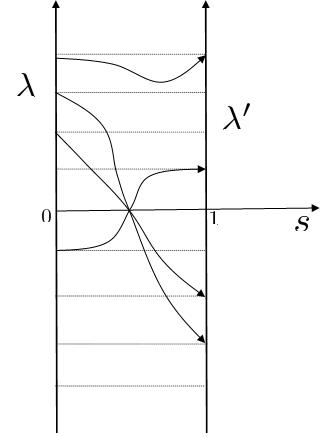}
\caption{When there are only three eigenvalues running across the
horizontal axis. Here only draw one particular diagram. There are
two lines goes from up to down and one from down to up, which
preserves $\nu_+-\nu_-=1$. Effectively, there are only one
eigenvalue reverse the sign, and all others mutually
rearranged.}\label{fig5}
\end{figure}
Then, when there are three eigenvalues flowing across the
horizontal axis, as shown in Fig.\ref{fig5}, similar analysis as
in the first case can tell there is a global gauge anomaly.

Finally, for any odd number of lines crossing the horizontal axis, it
is not difficult to draw the conclusion that for all the solutions
satisfying $\nu_+-\nu_-=1$, there is a global gauge anomaly. For any
even number of lines crossing the horizontal axis, there does not
exist a solution satisfying $\nu_+-\nu_-=1$. It ends the prove of
our equivalence property.

Up to now, we can conclude that the non-trivial topological quantum
spin Hall effect corresponds to the global gauge anomaly whose
gauge field is induced by the Berry connection of the former.
Here, a comment on global gauge anomaly may be helpful. In a
paper\cite{Alwis}, the author suggested that an $SU(2)$ global gauge
anomaly can be seen as an $U(1)$ anomaly by observing that the
special $U(1)$ element $e^{i\pi}$ is also the center of an $SU(2)$
transformation. So an $SU(2)$ global gauge anomaly can also be viewed
as a more familiar $U(1)$ anomaly as in the case of a $3D$ topological
insulator. However, here, the gauge group we are discussing is
more complicated than an $SU(2)$. But the existence of the above
correspondence may help us understanding our formulation by
studying the $3D$ case.

\section{Morse theory and the classical spin Hall effect}
\label{sec5}

In this section, we discuss the Witten-deformed Dirac
index\cite{Witten2} and find that $\mathbb{Z}_2$ index for quantum
theory is the same as that for the classical theory, and therefore
we conjecture that the classical "spin" Hall effect also contains the
$\mathbb{Z}_2$ topological property, which have not been discussed
as far as we know.

In the above, we have shown a crucial formula,
\begin{eqnarray}
\nu_+-\nu_-=\frac{1}{2\pi}\oint_{BZ} \textrm{tr}A, \label{5.1
index}
\end{eqnarray}
which is a specific example of Atiyah-Singer index theorem. In
fact, the left and right hand side of the equation are two
different limits of a same quantity\cite{Friedan}, $i.e.$
\begin{eqnarray}
\textrm{ind}(\slashed{D}[A])=\textrm{tr} \gamma_5 e^{-\beta \slashed{D}^{\dagger}\slashed{D}}=
\left\{
  \begin{array}{ll}
    \nu_+-\nu_-, & \hbox{$\beta\rightarrow\infty$;} \\
    \\
    \frac{1}{2\pi}\oint_{BZ} \textrm{tr} A, & \hbox{$\beta \rightarrow 0$.}
  \end{array}
\right. \label{5.2 index}
\end{eqnarray}
The index for Dirac operator reminds us the similar supersymmetric
case
\begin{eqnarray}
\textrm{Tr} (-1)^F e^{-\beta Q^{\dagger}Q} \label{5.3}
\end{eqnarray}
where $Q$ is the supersymmetry operator. These two index can be related by the following map:
\begin{eqnarray}
  (-1)^F &\rightarrow& \gamma_5, \\
  \slashed{D} &\rightarrow& Q, \\
  H=\slashed{D}^{\dagger}\slashed{D} &\rightarrow& H=Q^{\dagger}Q.
  \label{5.4-6}
\end{eqnarray}
E. Witten, in his seminal paper, derived the Morse inequality with
the help of supersymmetric quantum mechanics of the later case.
Here we apply his idea to the discussion of the Dirac index. By
parallel analysis, we can also derive the Morse inequality.

For convenience, we set the representation of the Dirac matrix
satisfying $\gamma_5= \textrm{diag}(I, -I)$. The Dirac operator
can be expressed as the form
\begin{eqnarray}
\slashed{D}=\left(
              \begin{array}{cc}
                0 & \slashed{D}_- \\
                \slashed{D}_+ & 0 \\
              \end{array}
            \right), \label{5.7}
\end{eqnarray}
where $\slashed{D}_-^{\dagger} = - \slashed{D}_+$, and $\slashed{D}_{\pm}=D_i\Gamma^i_{\pm}$. In the previous analysis, we focus on
the Dirac equation
\begin{equation}
i\slashed{D}_+\psi_+=\lambda\psi_+. \label{5.8}
\end{equation}
And other part is similar, $i.e.$
\begin{equation}
i\slashed{D}_-\psi_-=-\lambda\psi_-. \label{5.9}
\end{equation}
To counting the number of Dirac zero modes which is of interest in
the previous section, one only needs to count the chiral part or
anti-chiral part. Here, we will mainly focus on the ground state
of the Hamiltonian
\begin{eqnarray}
H&&=\textrm{tr}\slashed{D}^{\dagger}\slashed{D}\nonumber\\
&&=-(\slashed{D}_-\slashed{D}_++\slashed{D}_+\slashed{D}_-)\nonumber\\
&&=-2D^iD_i. \label{5.10}
\end{eqnarray}
We can show that the number of ground state of the Hamiltonian
equals to the number of the zero modes of the Dirac equation. If a
zero mode of a Dirac equation $\slashed{D}_+ \psi = 0$ with
$\slashed{D}_-\psi=0$ is simultaneously true, it is obvious that
$\psi$ is also the ground state of the Hamiltonian $H$. Conversely, if
$H\psi=0$, by integrating by parts, we can obtain
\begin{equation}
\int \psi^* H\psi=-2\int \psi^* D_iD^i\psi=2\int |D_i\psi|^2=0,
\label{5.12}
\end{equation}
which immediately leads to $D_i\psi=0$. Thus, $\psi$ is also the
zero mode of Dirac operator. Therefore, it is proven that we may
consider the ground states or the vacuum of the Hamiltonian instead of
counting the number of the Dirac zero modes. We should point out here
that this equivalence is only valid in the two dimensional case.

According to Witten's idea, we can deform the Dirac operator
slightly as the following
\begin{equation}
\slashed{D}_+\rightarrow\slashed{D}_+^t=e^{-ht}\slashed{D}_+e^{ht},
\label{5.13}
\end{equation}
\begin{equation}
\slashed{D}_-\rightarrow\slashed{D}_-^t=e^{ht}\slashed{D}_-e^{-ht}.
\label{5.14}
\end{equation}
Of course, deformations in this way neither introduce extra zero
ground states nor eliminate any. We may express the deformed
Dirac operator by introducing two more physical raising and
lowering operators
\begin{equation}
\slashed{D}_+^t=\slashed{D}_+ +tD_{i+}ha^i, \label{5.15}
\end{equation}
\begin{equation}
\slashed{D}_-^t=\slashed{D}_-+tD_{i-}ha^{i*}. \label{5.16}
\end{equation}
Here we may regard $a^i$ and $a^{j*}$ as rising and lowering one
chirality of the following operator respectively. Substituting above
into the Hamiltonian $H$, one can rewrite the Hamiltonian as the following
form
\begin{equation}
H=-\slashed{D}_-\slashed{D}_+-\slashed{D}_+\slashed{D}_-+t^2(iD_ih)^2+tiD_ih\
iD_jh[a^i, a^{j*}], \label{5.17 dH}
\end{equation}
where we let $i$ be explicit because the $i D$ is Hermitian and
the product with its conjugate is positive definited. One can
immediately see from the equation that the Hamiltonian is composed
of three parts, the kinetic term, the oscillator potential term,
and the quantum correction. The first two terms are purely
classical contributions. In order to extract out the classical
parts, it is natural to consider large $t$ limit under which case
the quantum correction, proportional to $t$, is suppressed down by
the quadratic of $t$. Under this limit, we may expand the eigenvalue
$\lambda$ of the total Hamiltonian by the order of $1/t$ as
\begin{equation}
\lambda(t)=t \left[A+\frac{B}{t}+\frac{C}{t^2}+\cdots\right].
\label{5.18}
\end{equation}
The $t$ in the front of Eq.(\ref{5.18}) is analogous to the
solution of the free oscillator which has the form of $\omega (n+1/2)$
where $\omega$ in this situation is $t$. $A$ represents the classical
contribution and other terms are the quantum effects.

For the ground state of the Hamiltonian, corresponding to
$\lambda(t)=0$, it implies that $A$, the energy eigenvalue,
corresponding the classical system, reaches its vacuum. However, the
reverse case is not necessarily true. Namely, for $A=0$, one cannot
obtain that the total quantum eigenvalue is zero. With this
consideration, one can write down an inequality
\begin{equation}
\#\{\lambda_p=0\}\leq \#\{A_p=0\}, \label{5.19}
\end{equation}
where $p$ means the chirality of the particle and the inequality
is valid for any $p$. It means that the number of quantum vacuum
is always less than that of the classical vacuum. Actually, as
Witten claimed, the number of vacuums on both sides can be
interpreted as the Betti-number and the Morse-index, respectively. The
argument will not be discussed here. The above inequality can be
rewritten as
\begin{equation}
B_p\leq M_p, \label{5.20}
\end{equation}
which is the weak form of the Morse inequality.

There is one more fact that the Morse inequality has a strong form, saying
\begin{equation}
\sum_p M_pt^p-\sum_pB_pt^p=(t+1)\sum_pQ_pt^p, \label{5.21}
\end{equation}
where $Q_p$'s are non-negative integers. Since, in the previous
section, we only count the total number of zero modes of the Dirac
operator, we should sum over all possible chirality. This
motivates us to set $t=1$. For $t=1$, the equality (\ref{5.21})
becomes
\begin{equation}
\sum_p M_p-\sum_pB_p=2\sum_pQ_p, \label{5.22}
\end{equation}
which gives us a great surprise. Since $B_p$, which is the number
of quantum vacuum, is stable only up to $\mathbb{Z}_2$, and the
right hand side of Eq.(\ref{5.22}) is always an even integer, so
the number of classical vacuum is stable up to $\mathbb{Z}_2$
also. Paraphrasing it into the condensed matter physics language,
the classical "spin" Hall effect also contains the $\mathbb{Z}_2$
topological invariant which characterizes its topological nature.
It is worth to note that in classical theory, there is no concepts
like spin, so we put a quotation mark on the spin. It is still
difficult to construct this classical system for the time being.
The method we try to use is to construct some kind of index which
counts the number of states satisfying $\langle k,\alpha|
\Theta|k,\beta\rangle=0$. Then, usually such index can be written
as some path integral. Then classical version of this is to use
the saddle point approximation. Then we can investigate such
quantity to see whether this is a $\mathbb{Z}_2$ quantity.
However, we are still not clear about how to construct the index.
We are trying to solve this and find other more accessible ways at
the same time. Moreover, this phenomenon is not covered in the
existing literatures, and more concrete analysis in terms of the
topological properties of the classical systems indicated by the
above is expected to be further investigated.

\section{Discussion and Conclusion}
\label{sec6}

In this article, we prove that the topological nontrivial quantum
spin Hall effect is equivalent to the $\mathbb{Z}_2$ global gauge
anomaly of the system described by the Lagrangian (\ref{4.9 L}). The
Lagrangian is somewhat similar to that describing massless QCD,
which also contains a pair of fermions and non-abelian
gauge fields. But the difference is that the former is defined on the
momentum space, and the fermions is a Weyl. Moreover, the
non-abelian gauge group, in order to generate a $\mathbb{Z}_2$
non-trivial topology, cannot be chosen arbitrarily. In fact, the
symmetry group is determined by the CPT symmetry of the
Hamiltonian. Similar to the fact that there exist fractional charge in QCD, the
fractional excitations in the QSHE also appear\cite{Lan}, which needs
further investigation.

As we mentioned before, the appearance of a global gauge anomaly
can occur in any even dimension, therefore, we expect that we may
find similar correspondence between a $4+1$ dimensional
topological insulator and a $4d$ non-abelian gauge
theory(Euclidean QCD). Moreover, from the approach of
Ref.\cite{QHZ}, we find the $2+1$ dimensional topological
invariant, which is obtained by a dimensional reduction, is
\begin{equation}
P_3(\theta)=\frac{1}{2\pi}\oint d\phi\Omega_{\phi},
\end{equation}
where $\Omega_{\phi}$ is defined in Eq.(112) in Ref.\cite{QHZ}. It
is of the same form as (\ref{4.7 I}). Therefore, we can conclude
that our formalism can also be applied to a Chern-Simons field
theory approach. However, we should mark that the $3+1$
dimensional case, which is not included in our formalism, is
related to the $U(1)$ chiral anomaly. Other $\mathbb{Z}$ SPT cases
were already found to be corresponding to various anomalies. The
relationship between an anomaly and most of other $\mathbb{Z}_2$
topological SPT states is still not clear yet. We hope that our
article can provide some inspiration toward the understanding of
these issues.

Lastly, through some analysis of the spectrum of the deformed
Hamiltonian (\ref{5.17 dH}), we find that the classical "spin" Hall
effect also has a $\mathbb{Z}_2$ topological classification, which
is not discovered before as far as we know. By further investigation of a
classical system, we may find more interesting topological nature
which may enable us a better understanding of the topology of
strong correlated systems.

\section*{Acknowledgement}
The work was supported by National Natural Science Foundation of
China under Grant No.11275180 and National Science Fund for
Fostering Talents in Basic Science No.J1103207.

\end{document}